\documentclass[aps,pra,onecolumn,preprint,preprintnumbers,groupaddress,amsmath,amssymb,usenatbib,
nofootinbib,showkeys,showpacs,altaffilletter,11pt]{revtex4}
\usepackage{graphicx}
\usepackage{dcolumn}
\usepackage{bm}
\usepackage{latexsym}
\usepackage{amsfonts}
\usepackage{amssymb}
\usepackage{amsmath}
\usepackage{subfigure}
\usepackage{mathrsfs}
\usepackage{tabu}
\usepackage{xcolor}
\usepackage{soul}
\usepackage[colorlinks]{hyperref}

\begin{document}

\title{Thermodynamics and stability of $f(T,B)$ gravity with viscous fluid by observational constraints}

\author{A. Pourbagher}
\email{st.a.pourbagher@iauamol.ac.ir}
\author{Alireza Amani}
\email{a.r.amani@iauamol.ac.ir}
\affiliation{\centerline{Department of Physics, Ayatollah Amoli Branch, Islamic Azad University,} \\Amol, Mazandaran, Iran}

\date{\today}

\begin{abstract}
In this paper, we study the model of $f(T, B)$ gravity with viscous fluid in flat-FRW metric, in which $T$ and $B$ are torsion scalar and boundary term, respectively. We obtain the Friedmann equations in the framework of modified teleparallel gravity by tetrad components. We consider an interacting model between matter and dark energy so that universe dominates by viscous fluid. Then, we write the corresponding cosmological parameters in terms of the redshift parameter, and, we parameterize the Hubble parameter with experimental data. In what follows, we plot the corresponding cosmological parameters for dark energy components in terms of redshift, thereafter we investigate the accelerated expansion of the universe. Moreover, we discuss the stability of the model by using the sound speed parameter. Finally, we investigate the validity of the generalized second law of thermodynamics.
\end{abstract}
\maketitle

\section{Introduction}\label{I}

In the last two decades, the accelerated expansion of the universe was first discovered in type Ia supernova data, and it reconfirmed by cosmic microwave background and large scale structure \cite{Riess_1998, Perlmutter_1999, Bennett_2003, Tegmark_2004}. The result of these efforts led to the introduction of mysterious energy called dark energy, which is now one of the great challenges of general relativity. This means that dark energy is a hypothetical form of energy that is driving the accelerated expansion of the universe, so, it is essential that the universe is in a phase of negative pressure, and one is three-quarters of the total energy of the universe. Over the past two decades, efforts have been made to understand the cause of acceleration and introduce models to describe it. For this purpose, extensive studies have been carried out by many researchers on various dark energy models in isotropic space-time such as cosmological constant, scalar fields, tachyon field, modified gravity models, interacting models, bouncing model and braneworld models \cite{Weinberg-1989, Caldwell-2002, Amani-2011, Sadeghi1-2009, Setare-2009, Setare1-2009, Battye-2016, Li-2012, Sadeghi-2013, Sadeghi-2014, Pourhassan-2014, JSadeghi-2015, Amani1-2015, Iorio-2016, Faraoni-2016, Khurshudyan1-2014, Sadeghi1-2016, Wei-2009, Amani1-2011, Amani-2015, Nojiri_2006, Nojiri_2007, Li-2004, Campo-2011, Hu-2015, Fayaz-2015, Saadat1-2013, Amani-2013, Amani-2014, Naji-2014, KhurshudyanJ-2014, KhurshudyanB-2014, Morais-2017, Zhang1-2017, Bouhmadi1-2016, KhurshudyanJ2-2014, SadeghiKhurshudyanJ1-2014, Sadeghi-2010, Sadeghi-2009, Amani-2016, Singh-2016, Sahni1-2003, Setare-2008, Brito-2015, Setaremr-2008, Amanifarahani-2012, Amanifarahani1-2012, Amanipourhassan-2012, Chirde-2016, Bhoyar-2017, Chirde-2018, Chirde1-2018}. Some other models can be referred to as modified teleparallel models \cite{Linder_2010, Myrzakulov_2011, Li_2011, Myrzakulov_2012, Harko_2011, Nojiri_2005}. It should be noted that the solution of the aforesaid models provides an appropriate alternative for describing the accelerated expansion of the universe.

As we know, the Einstein-Hilbert action is written in terms of the curvature term, $R$, where describes gravity in general relativity. The theory of general relativity is very successful in agreement with observation data. In addition, another gravitational scenario has been proposed so-called teleparallel gravity, which was first introduced by Einstein \cite{Einstein_1928}. Thus, the mathematical structure of distant parallelism was introduced for the unification of electromagnetism and gravity by tetrad components. We note that tetrad components are derived with the Weitzenb\"{o}ck connection in teleparallelism and metric components are derived with the Levi-Civita connection in the framework of general relativity. Therefore, teleparallel gravity scenario is replaced instead of general relativity by using the transformation of the tetrad components to the metric components \cite{Weitzenbock_1923, Bengochea_2009}. In other words, the curvature term $R$ in general relativity is transformed to a torsion term $T$ in the teleparallel scenario, which modified form changes from $T$ to $f(T)$ by an arbitrary function in the corresponding action so-called $f(T)$ cosmology theory \cite{Cardone_2011, Jamil_2012, Sharif_2012, Shekh_2019, Mirza_2019}.

Some recent models of teleparallel gravity can be introduced as scenarios of  $f(T,\mathcal{T})$ gravity and $f(T, B)$ gravity, in which $\mathcal{T}$ and $B$ are a trace of the matter energy-momentum tensor and boundary term, respectively \cite{Harko_2014, Rezaei_2017, Bahamonde1_2017}. It is interesting to note that these models have good compatibility with observational data to describe the accelerated expansion of the universe. In particular, the $f(T,B)$ gravity has a better advantage because one recovers the same formation in both scenarios of $f(T)$ gravity and $f(R)$ gravity, which is resulted equivalence teleparallel--curvature and so-called $f (R,T)$ gravity. Therefore, the $f(T,B)$ gravity is another form of $f (R,T)$ gravity, which is a great alternative to dark energy \cite{Amani1-2015, Amani-2015, Myrzakulov_2012}.

In addition to the aforesaid theories, the existence of anisotropic fluid is another factor in the variation of cosmic acceleration. The most important models of anisotropic fluid are Bianchi model and viscous fluid model \cite{Sadeghi_2013, Pourhassan-2013, SaadatmBB-2013, Amaniali-2013}. For this purpose, we will consider viscous fluid in this job, namely, that the universe is dominated by the viscous fluid. So, this issue gives us the motivation to investigate the dynamical effects of bulk viscosity on the $f(T, B)$ gravity.

Now, we are going to study the $f(T, B)$ gravity in the presence of bulk viscosity by interacting between components of matter and dark energy. Then, the Hubble parameter will fit by an algebraic function with the astronomical data. Since there is a fundamental connection between gravitation and black hole thermodynamics, then the connection can be created between horizon entropy and the area of a black hole \cite{Bardeen-1973, Hawking-1975, Zubair-2017}. Inspired by this view, we will examine the validity of our model using the second law of thermodynamics.

This paper is organized as the following:\\
In Sec. \ref{II}, we implement the fundamental formation of $f(T,B)$ gravity in the flat-FRW metric. In Sec. \ref{III}, we consider that universe dominates by a bulk viscosity fluid, and also obtain the Friedmann equations and equation of state (EoS). In Sec. \ref{IV}, we parameterize our model by redshift parameter, and then the Hubble parameter is drawn by fitting to astronomical data, and also the stability of the model be investigated by the function of sound speed. In Sec. \ref{V} we investigate the generalized second law of thermodynamics for our model validation. Finally, in Sec. \ref{VI} we will give result and conclusion for the model.

\section{The $f(T,B)$ gravity}\label{II}
In this job, we intent to consider a combine theory by $f(R)$ gravity and $f(T)$ gravity \cite{Bahamonde-2015, Bahamonde-2017} by action
\begin{equation}\label{act1}
S=\int d^4x \, e \, \left(\frac{f(T,B)}{\kappa^2} \, +\mathcal{L}_m\right),
\end{equation}
where $\kappa^2 = 8 \pi G$, and $f(T, B)$  is an arbitrary function of  the tortion scalar $T$ and the boundary term $B=\frac{2}{e} \partial_\mu \left(e T^\mu \right)$ in which $T_\mu=T^\nu _{\nu \mu}$. The $\mathcal{L}_m$ and $e=det(e^i_\mu)$ are matter action and determinant of tetrad components, respectively. The relationship between tetrad and metric is written as $e^i_\mu$$=\sqrt{-g}$ and $g_{\mu \nu}=\eta_{i j} e^i_\mu e^j_\nu$  in which $g$  and $\eta_{i j}=diag(1,-1,-1,-1)$ are determinant of metric components $g_{\mu \nu}$ and the Minkowski metric tensor, so that indices of Greek and Latin are space--time and tangent space--time, respectively. However, the tensor relationships are written by
\begin{subequations}\label{TT1}
\begin{eqnarray}
 &T^{\lambda}_{\mu\nu}=\overset{\mathbf{w}}\Gamma^{\lambda}_{\nu\mu}-\overset{\mathbf{w}}\Gamma^{\lambda}_{\mu\nu}=e^{\lambda}_i(\partial_\mu e^i_{\nu}-\partial_\nu e^i_{\mu}),\label{TT1-1}\\
 &T=\frac{1}{4}T^{\rho\mu\nu}T_{\rho\mu\nu}+\frac{1}{2}T^{\rho\mu\nu}T_{\nu\mu\rho}-T^{\rho}_{\rho\mu}T^{\nu\mu}_{\nu},\label{TT1-2}\\
 &S_{\rho }^{\mu \nu } =\frac{1}{2}\left( {k}^{\mu \nu }_{\rho }+\delta _{\rho }^{\mu } \, {T}^{\alpha \nu }_{\alpha }-\delta _{\rho }^{\nu } \, {T}^{\alpha \mu }_{\alpha } \right),\label{TT1-3}\\
 &{k}^{\mu \nu }_{\rho }=-\frac{1}{2}\left( {T}^{\mu \nu }_{\rho }+{T}^{\nu \mu }_{\rho }-{T}_{\rho }^{\mu \nu } \right),\label{TT1-4}
\end{eqnarray}
\end{subequations}
where $\overset{\mathbf{w}}\Gamma^{\lambda}_{\mu\nu}= e^{\lambda}_i \partial_{\mu} e^{i}_{\nu}$, $T^{\lambda}_{\mu\nu}$, $S_{\rho }^{\mu \nu }$  and ${k}^{\mu \nu }_{\rho }$ are the Weitzenb\"ock connection, the torsion tensor, the asymmetry tensor and  the contorsion tensor, respectively.

The following field equation obtains by varying the action \eqref{act1} with respect to tetrad as
\begin{eqnarray}\label{frid1}
\begin{aligned}
2e\delta _\nu ^\lambda {\nabla ^\mu }{\nabla _\mu }{\partial _B}f - 2e{\nabla ^\lambda }{\nabla _\nu }{\partial _B}f + eB{\partial _B}f\delta _\nu ^\lambda  + 4e\left( {{\partial _\mu }{\partial _B}f + {\partial _\mu }{\partial _T}f} \right){S_\nu }^{\mu \lambda } + 4e_\nu ^a{\partial _\mu }\left( {e{S_a}^{\mu \lambda }} \right){\partial _T}f\\
- 4e{\partial _T}f{T^\sigma }_{\mu \nu }{S_\sigma }^{\lambda \mu } - ef\delta _\nu ^\lambda  = 16\pi e\mathcal{T}_\nu ^\lambda.
\end{aligned}
\end{eqnarray}

According to the cosmology principle, the universe has both homogeneous and isotropic space-time models. However, here we consider the flat metric of FRW in the following form
\begin{equation}\label{ds21}
ds^2=dt^2-a^2(t)\left(dx^2+dy^2+dz^2 \right),
\end{equation}
where $a(t)$ is the scale factor. Then, the vierbein field and the torsion scalar are written by metric \eqref{ds21} as
\begin{subequations}\label{eimuT1}
\begin{eqnarray}
 &e^i_\mu = e^\mu_i = diag(1, a, a, a),\label{eimuT1-1}\\
 &T=-6 H^2.\label{eimuT1-2}
\end{eqnarray}
\end{subequations}

The important point of the job is that both the torsion scalar and Ricci scalar are related together as
\begin{equation}\label{rtb}
R = -T + B,
\end{equation}
this means that standard action of general relativity is made with Ricci scalar $R$, but the $f(T,B)$ action is made with torsion scalar and boundary term. This issue tell us that these models are different only by a boundary term (for more details see Refs. \cite{Bahamonde-2017, Bahamonde-2018}). The boundary term obtains by metric \eqref{ds21} in the following form
 \begin{equation}\label{rtb1}
B = -6\left (\dot{H}+3H^2\right),
\end{equation}
then the Ricci scalar $R$ is re-found from \eqref{rtb} as $R= -6\left (\dot{H}+2 H^2\right)$.

\section{Viscous fluid model }\label{III}
As we know, in standard cosmology model, the universe  is considered by a perfect fluid, but in a realistic theory we consider that the universe is content a non-perfect fluid. This issue tells us that there is a bulk viscosity in the depth of cosmic. Therefore, the its energy--momentum tensor is written  as
\begin{equation}\label{Tij1}
\mathcal{T}_i^j=(\rho + p + p_{b})u_i u^j - \left(p + p_{b}\right)\,  \delta_i^j,
\end{equation}
where the energy density $\rho$ and the pressure $p$ are the cosmological parameters of fluid inside the universe, and $p_{b} = -3 \xi H$ represents the bulk viscosity pressure in which $\xi$ is introduced as positive constant coefficient of bulk viscosity. The $4$-velocity $u_\mu$ is $u^i$ = (+1,0,0,0) and can be written $ u_i u^j$ = 1. However, we can write the non-zero elements of energy--momentum tensor by
\begin{equation}\label{tau1}
 \mathcal{T}_0^0 = \rho, \\
 \mathcal{T}_1^1 = \mathcal{T}_2^2 = \mathcal{T}_3^3 = -p + 3 \xi H.
\end{equation}
From the Eq. \eqref{frid1} we can obtain the field equations as
 \begin{subequations}\label{freid2}
\begin{eqnarray}
& - 3{H^2}\left( {3{\partial _B}f + 2{\partial _T}f} \right) + 3H{\partial _B}\dot f - 3\dot H{\partial _B}f + \frac{1}{2}f = {\kappa ^2} \rho,\\
& - \left( {3{H^2} + \dot H} \right)\left( {3{\partial _B}f + 2{\partial _T}f} \right) - 2H{\partial _T}\dot f + {\partial _B}\ddot f + \frac{1}{2}f =   -{\kappa ^2} (p - 3 \xi H),
\end{eqnarray}
\end{subequations}
where the dot symbol is derivative with respect to time. However, we can write the Friedmann standard form as
 \begin{subequations}\label{freidstan1}
\begin{eqnarray}
 &3{H^2}=\kappa^2 \rho_{tot},\label{freidstan1-1}\\
  &2 \dot H +3{H^2}=  -{\kappa ^2} \bar{p}_{tot},\label{freidstan1-2}
\end{eqnarray}
\end{subequations}
where $\bar{p}_{tot}=p_{tot} - 3 \xi H$, in which $p_{tot}$ is the total pressure of the universe that is contained with the matter and dark energy pressures.  In what follows, the corresponding continuity equation for perfect fluid is written by
\begin{equation}\label{contequ1}
\dot{\rho}_{tot}+3 H \left(\rho_{tot} + \bar{p}_{tot}\right)=0.
\end{equation}
Also, we introduce the EoS of the whole Universe despite the viscosity in the following form
\begin{equation}\label{eostde1}
\omega_{tot} = \frac{\bar{p}_{tot}}{\rho_{tot}}.
\end{equation}

Now here we investigate an interacting model for the current job. So, we consider that the Universe dominates by viscous fluid and $f(T,B)$ gravity. In that case, the parameters $\rho_{tot}$ and  $p_{tot}$ are written as follows:
\begin{subequations}\label{rhopresstot1}
\begin{eqnarray}
 &\rho_{tot} = \rho + \rho_d,\label{rhopresstot1-1}\\
 &p_{tot} = p + p_d,\label{rhopresstot1-2}
\end{eqnarray}
\end{subequations}
where the quantities $\rho_d$ and $p_d$ come from $f(T,B)$ gravity and these are the causes of dark energy. Then, by using the Eqs. \eqref{freid2}, \eqref{freidstan1} and \eqref{rhopresstot1} we find $\rho_d$ and $p_d$ as
\begin{subequations}\label{rhod1}
\begin{eqnarray}
 &\kappa^2 \rho_{d} = 3 H^2 \left( 1+3 \partial_B f +2 \partial_T f \right) - 3 H \partial_B \dot{f} +3 \dot{H} \partial_B f -\frac{1}{2} f,\label{rhod1-1}\\
 &\kappa^2 p_{d} =  - {3{H^2}}\left( {1 + 3{\partial _B}f + 2{\partial _T}f} \right) - \dot{H}\left( {2 + 3{\partial _B}f + 2{\partial _T}f} \right) - 2H{\partial _T}\dot f + {\partial _B}\ddot f + \frac{1}{2}f,\label{rhod1-2}
\end{eqnarray}
\end{subequations}
so, the Eqs. \eqref{rhod1} shows us that the density energy and pressure of dark energy depend on $f(T, B)$ gravity. Now we consider an interacting model between components of universe, this means that there is an energy flow between both components. In that case, the continuity equation \eqref{contequ1} is separately written  for components of matter and dark energy in the following form
\begin{subequations}\label{contin2}
\begin{eqnarray}
 &\dot{\rho}+3 H \left(\rho + p \right)= Q,\label{contin2-1}\\
 &\dot{\rho}_{d}+3 H \left(\rho_{d} + \bar{p}_{d}\right)= -Q,\label{contin2-2}
\end{eqnarray}
\end{subequations}
where $\bar{p}_d = p_d - 3 \xi H$, and $Q$ is interaction term between the matter and dark energy, so that, when $Q>0$, energy flow transfer from dark energy to the matter, and when $Q<0$, energy flow transfer vice versa. We here consider interaction term as $Q = 3 b^2 H \rho$ in which $b$ is the transfer strength or coupling parameter. The solution of first order differential equation \eqref{contin2-1} obtains as
\begin{equation}\label{rhom}
\rho = \rho_0 \,a^{-3 (1-b^2+\omega)},
\end{equation}
where $\omega=\frac{p}{\rho}$ is the EoS of matter, and $\rho_0$ is the matter energy density of current universe. The EoS of dark energy is written by
\begin{equation}\label{eos1}
\omega_d = \frac{\bar{p}_d}{\rho_d},
\end{equation}
we note that the corresponding EoS is dependent on parameters of $f(T,B)$ gravity and viscous fluid.

\section{Parameterization and stability analysis}\label{IV}

In this section, we are going to reconstruct the $f(T,B)$ gravity model with redshift parameter. To achieve this goal, we try to write some cosmological parameters such as the energy density and the pressure of dark energy with respect to redshift parameter. As we know, the scale factor is written as  a function of redshift parameter $z$ as $1+z=\frac{a_0}{a}$ in which $a_0$ is the late time scale factor. The relationship between the time derivative and the  redshift derivative is $\frac{d}{dt} = -H(1+z)\frac{d}{dz}$. The first, we re-write parameters, $T$, $B$ and $R$ versus redshift parameter as
\begin{subequations}\label{tbr1}
\begin{eqnarray}
 &T = -6 H_0^2 E(z),\label{tbr1-1}\\
 &B =  3 H_0^2 (1+z) E'(z)-18 H_0^2 E(z), \label{tbr1-2}\\
 &R =  3 H_0^2 (1+z) E'(z)-12 H_0^2 E(z),\label{tbr1-3}
\end{eqnarray}
\end{subequations}
where $H^2 = H_0^2 \,E(z)$ in which $H_0 = 68 \pm 2.8 \, km s^{-1} Mpc^{-1}$ is the late time Hubble parameter and the prime demonstrates the derivative with respect to redshift parameter. Then, from Eqs. \eqref{rhod1} we find the energy density and the pressure of dark energy in terms of redshift parameter as follows:
\begin{subequations}\label{rhod2}
\begin{eqnarray}
 &\rho_{d} =\frac{1}{\kappa^2} \Big[ 3 H_0^2 E \left( 1+3 \partial_B f +2 \partial_T f \right) + 3 H_0^2 (1+z) E \partial_B {f'} - \frac{3}{2} H_0^2 (1+z) E' \partial_B f -\frac{1}{2} f\Big],\label{rhod2-1}\\
 &\overline{p}_{d} =  \frac{1}{\kappa^2} \Big[- {3{H_0^2 E}}\left( {1 + 3{\partial _B}f + 2{\partial _T}f} \right) + \frac{1}{2} H_0^2 (1+z) E' \left( {2 + 3{\partial _B}f + 2{\partial _T}f} \right)  + 2 H_0^2 (1+z) E {\partial _T}f' \nonumber \\
& +H_0^2 (1+z)^2 E' {\partial _B}f'+ 2 H_0^2 (1+z) E {\partial _B}f' +H_0^2 (1+z)^2 E {\partial _B}f'' + \frac{1}{2}f \Big]-3 \xi H_0 \sqrt{E}.\label{rhod2-2}
\end{eqnarray}
\end{subequations}

Now we get a power-law like function by
\begin{equation}\label{ftb1}
f (T, B) = \alpha B^m + \beta T^n,
\end{equation}
where $\alpha$, $\beta$, $m$ and $n$ are constant.

\begin{figure}[t]
\begin{center}
\includegraphics[scale=.3]{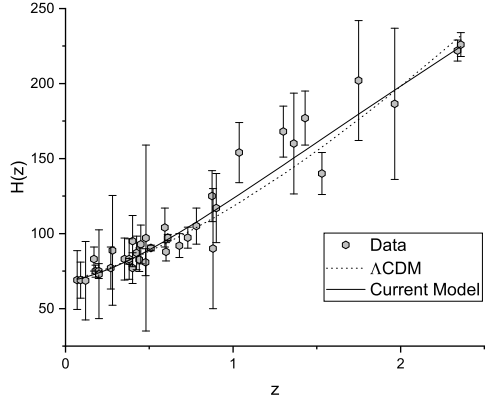}
\caption{The graph of the Hubble parameter in terms of redshift parameter for data \cite{Farooq_2017} (dot $+$ error bar), $\Lambda CDM$ model (dash) and our model (line).}\label{fig1}
\end{center}
\end{figure}

To solve the above equations and inspiration from Ref. \cite{Sahni-2003}, we consider the parametrization of function $E(z)$ in the following model,
\begin{equation}\label{rz1}
E(z) =A_3 (1+z)^3 + A_2 (1+z)^2 + A_1 (1+z) + A_0,
 \end{equation}
  where $A_3$, $A_2$, $A_1$ and $A_0$ are constant, and their measurements are determined by fitting the experimental data. Also, there is a constraint between these coefficients by condition $E(z=0) = 1$ as $A_3 + A_2 + A_1 +A_0 =1$. We set for the flat-$\Lambda CDM$ model by the parametrization $A_1 = A_2 =0$ and $A_0 = 1- A_3$, and one set by parametrization $A_2 =0$ and $A_0 = 1- A_1- A_3$ for the curve-$\Lambda CDM$ model. In what follows, we want to fit the parametrization function \eqref{rz1} by $38$ supernova data of  Ref. \cite{Farooq_2017} in which Hubble parameters were measured in terms of redshift parameter, notice that the data were collected from Refs. \cite{Simon_2005, Stern_2010, Moresco_2012, Blake_2012, Font_2014, Delubac_2015, Moresco_2015, Alam_2016, Moresco_2016}. From the result of fitting, we obtain the coefficients of the parametrization function \eqref{rz1} as $A_3 = -0.16 \pm 0.30$, $A_2 =  2.39 \pm 1.71$, $A_1= -3.80 \pm 2.89$ and $A_0 = 1-A_3-A_2-A_1$ which Fig. \ref{fig1} shows the matter.

\begin{figure}[h]
\begin{center}
\includegraphics[scale=.3]{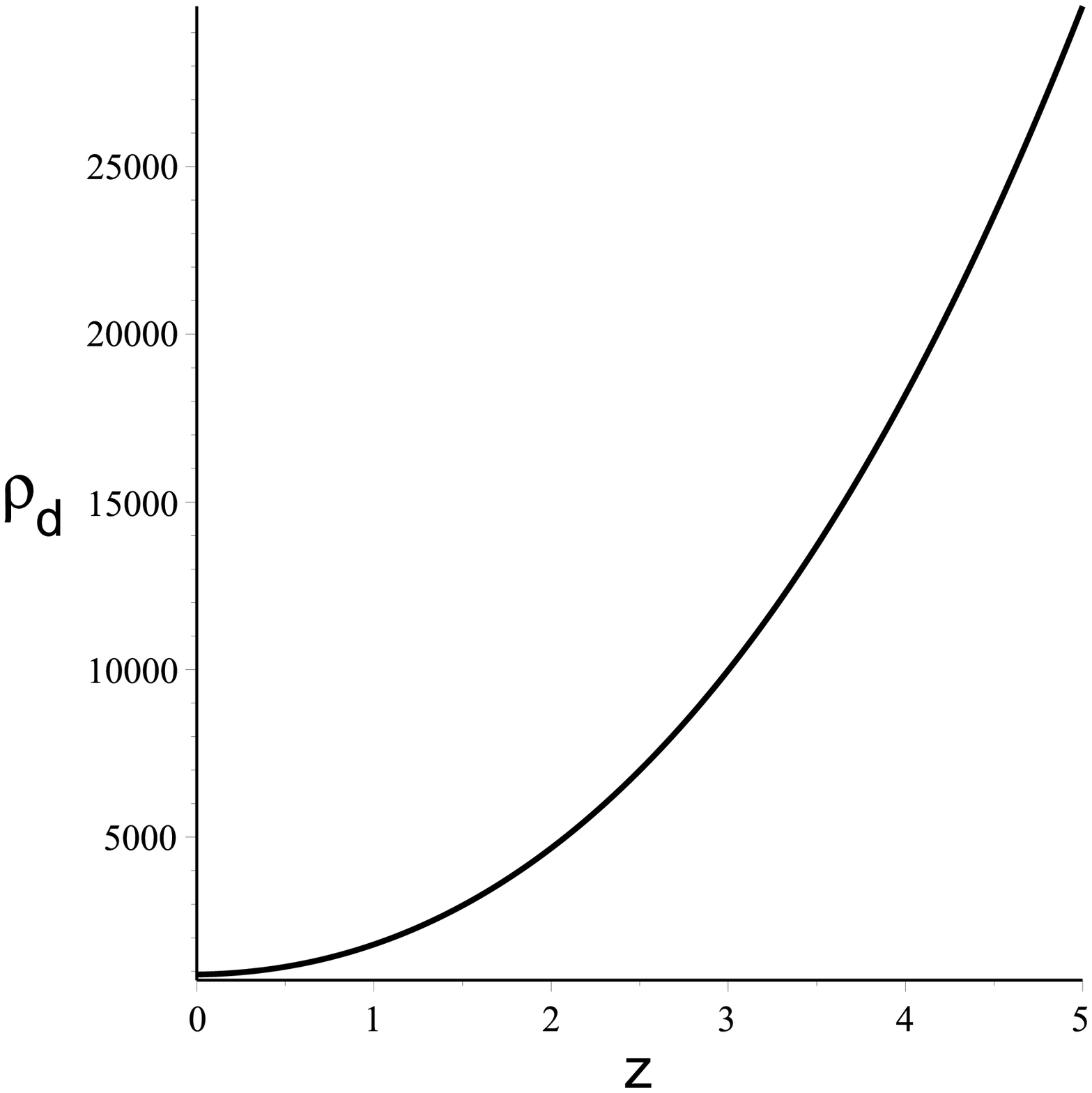} \includegraphics[scale=.3]{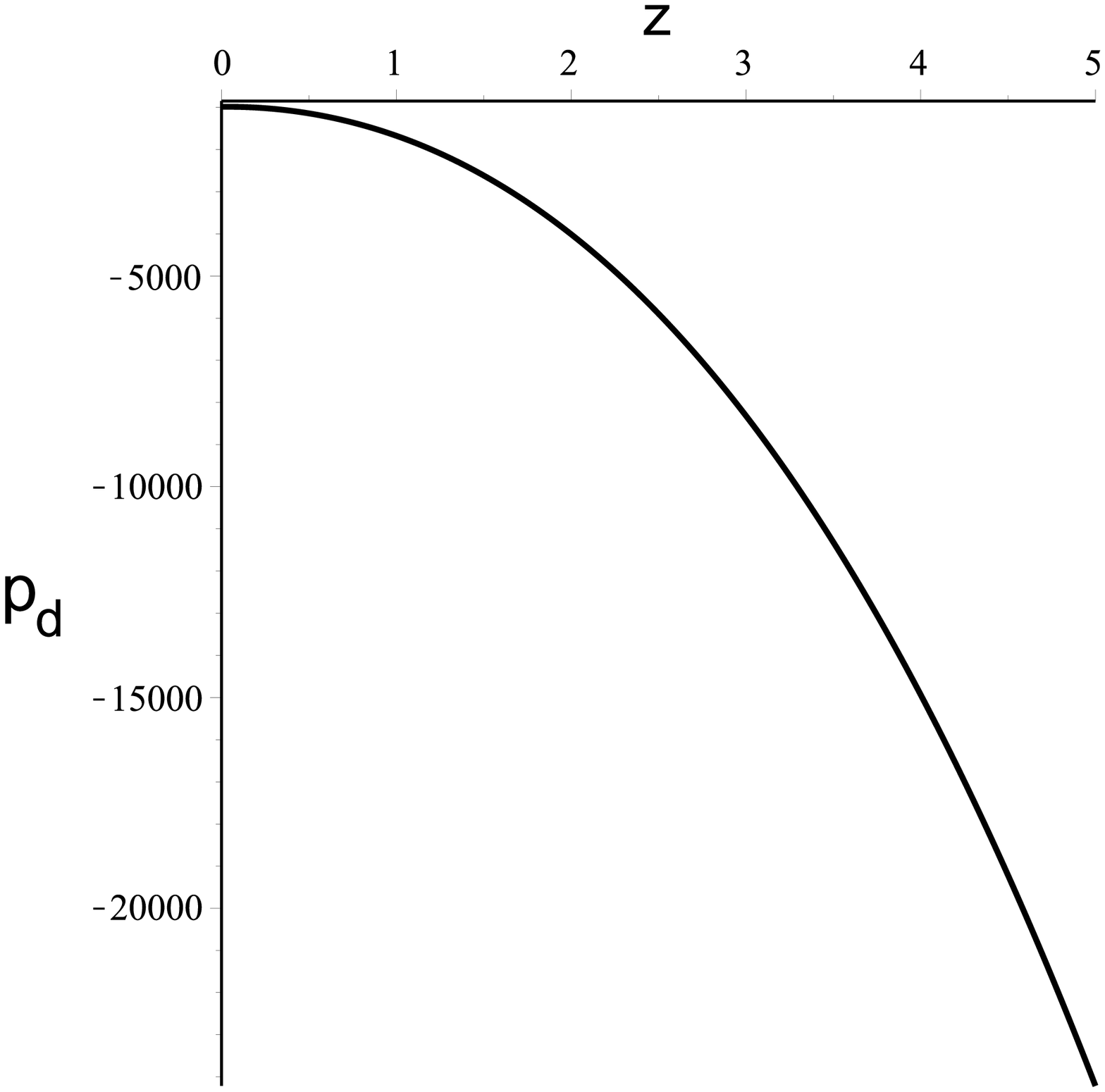}
\caption{The graph of the energy density and the pressure of dark energy in terms of redshift parameter.}\label{fig2}
\end{center}
\end{figure}

  In that case, by inserting Eqs. \eqref{tbr1-1} and \eqref{tbr1-2} into Eq. \eqref{ftb1}, and then by substituting the obtained Eq. \eqref{ftb1} and Eq. \eqref{rz1} into Eq. \eqref{rhod2}, we can draw the graph of energy density and pressure of dark energy in terms of redshift parameter in Fig \ref{fig2}. Here we note that the free parameters play an very important role in our results, so we try to choose them as $\alpha = \beta = 1$, $ m = 1$, $ n =  -2$, $\xi  =  0.5$, $ A_3 =  0.14$, $ A_2 =  0.75$, $A_1 =  -1.45$ and $A_0 = 1.56$. These coefficients are selected with the motivation that the energy density of dark energy is greater than zero and the pressure of dark energy is less than zero to confirm the accelerated expansion of the universe.

\begin{figure}[h]
\begin{center}
\includegraphics[scale=.3]{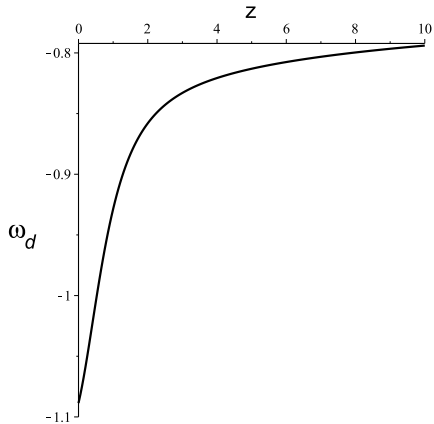}
\caption{The graph of the EoS of dark energy in terms of redshift parameter.}\label{fig3}
\end{center}
\end{figure}

By calculation the corresponding energy density and pressure of dark energy, we can plot the variations of EoS of dark energy \eqref{eos1} in terms of redshift variable as Fig. \ref{fig3}. This figure shows us crossing the phantom divided line, so that the current value of EoS ($z=0$) is equal to $-1.09$. Therefore, we conclude that the universe is undergoing an accelerated expansion and this issue is compatible with cosmological data of Ref. \cite{Amanullah_2010}.

\begin{figure}[h]
\begin{center}
\includegraphics[scale=.3]{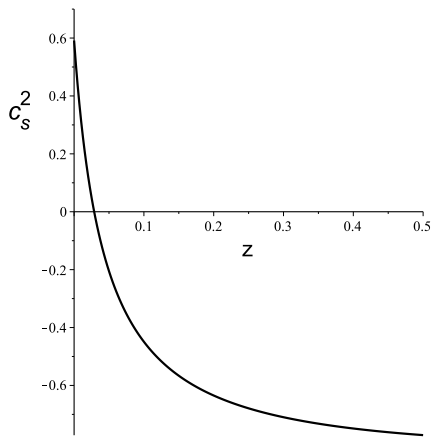}
\caption{The graph of the sound speed in terms of redshift parameter.}\label{fig4}
\end{center}
\end{figure}

In what follows, we consider the variation of dark energy pressure in terms of entropy and energy density in a general thermodynamics as $p_d = p_d (S,\rho_d)$. In that case, the pressure perturbation is written in a adiabatic system ($\delta S = 0$) by $\delta p_d = \frac{\partial p_d}{\partial \rho_d} \delta \rho_d$, in which $c_s^2=\frac{\partial p_d}{\partial \rho_d}=\frac{\partial_z p_d}{\partial_z \rho_d} $ is introduced as sound speed parameter, and the stability condition should be set to $c_s^2 > 0$. Therefore, we can see in the Fig. \ref{fig4} the value of $c_s^2$ in late-time ($z = 0$) with an amount of $0.59$, which satisfies stability condition. However, Fig.  \ref{fig4} shows us that there is a stability in late-time.

\section{Thermodynamics analysis}\label{V}

In this section, we intend to investigate the generalized second law of thermodynamics for the viscous $f(T,B)$ model in a flat--$FRW$ universe. In order to examine the validity of the second law of thermodynamics, we formulate the constraint for the model. The generalized second law of thermodynamic tell us that the entropy of inside horizon plus the entropy of boundary of horizon always increases versus time evolution \cite{Bekenstein_1973, Hawking_1975, Karami_2012}. In that case, radius of apparent horizon $r_h = a(t) r$ obtains in terms of the Hubble parameter by the condition $h^{\alpha \beta} \partial_\alpha r_h \partial_\beta r_h = 0$ for flat--$FRW$ metric as
\begin{equation}\label{rh}
  r_h = \frac{1}{H},
\end{equation}
where $\alpha, \beta = 0,1$ and the $2$-dimensional metric $h_{\alpha, \beta} = diag (1, -a^2)$. By taking the derivative from the above relation with respect to time one gets
\begin{equation}\label{rh1}
d r_h = - \frac{\dot{H}}{H^2} dt.
\end{equation}
We assume that the universe is in a condition of thermal equilibrium, in which case the internal temperature is same with the apparent horizontal temperature. The Hawking temperature or associated temperature is $T_h = \frac{\kappa_{sg}}{2 \pi}$, which $\kappa_{sg}$ is the surface gravity and is equal to
\begin{equation}\label{surgrav1}
\kappa_{sg} = \frac{1}{2 \sqrt{-h}} \partial_\alpha \left(\sqrt{-h} h^{\alpha \beta} \partial_b r_h\right) = \frac{1}{r_h}\left(1-\frac{\dot{r}_h}{2 H r_h}\right) = \frac{r_h}{2}\left(2 H^2 + \dot{H}\right).
\end{equation}

Now we can define Bekenstein--Hawking horizon entropy in general relativity by $S_{oh} = \frac{A}{4 G}$, in which $A = 4 \pi r_h^2$  is the area of the apparent horizon. Nonetheless, we clearly find
\begin{equation}\label{antropy1}
 T_h \dot{S}_{oh} = \frac{\dot{r}_h}{G}\left(1-\frac{\dot{r}_h}{2}\right).
\end{equation}

On the other hand, the Gibbs equation is written for the entropy of contents within the apparent horizon as
\begin{equation}\label{antropy2}
 T_h dS_{ih} = d(\rho_{tot} V) + \bar{p}_{tot} dV = V d\rho_{tot} + (\rho_{tot} + \bar{p}_{tot}) dV,
\end{equation}
where $\rho_{tot}$ and $\bar{p}_{tot}$ come from \eqref{freidstan1}, and $V =\frac{4}{3} \pi r^3_h$. To derivative of Eq. \eqref{antropy2} with respect to time, and by using the continuity equation \eqref{contequ1} yields
\begin{equation}\label{antropy3}
 T_h \dot{S}_{ih} = 4 \pi r_h^2 (\dot{r}_h-1) \left(\rho_{tot} + \bar{p}_{tot} \right).
\end{equation}

As we know, the generalized second law of thermodynamics expresses that the entropy of contents inside the apparent horizon plus the entropy on boundary of apparent horizon should not decreased. This means that the validity of the generalized second law of thermodynamics requires the condition
\begin{equation}\label{antropytot}
 \dot{S}_{tot} = \dot{S}_{ih}+\dot{S}_{oh}\geq 0,
\end{equation}
where by instituting Eqs. \eqref{antropy1} and \eqref{antropy3} into Eq. \eqref{antropytot} write the total entropy of universe as
\begin{equation}\label{antropytot1}
 T_h \dot{S}_{tot} = \frac{\dot{r}_h}{G}\left(1-\frac{\dot{r}_h}{2}\right)+4 \pi r_h^2 (\dot{r}_h-1) \left(\rho_{tot} + \bar{p}_{tot} \right),
\end{equation}
now by inserting \eqref{freidstan1} and \eqref{rh1} into aforesaid relationship, we can clearly find the condition for the validity of the generalized second law of thermodynamics in the following form
 \begin{equation}\label{antropytot2}
 \frac{\dot{H}^2}{2 G H^4}=\frac{(1+z)^2 E'^2(z)}{2 G E^2(z)}\geq 0.
\end{equation}

The result shows us that the validity of the generalized second law of thermodynamics be satisfied by condition of thermodynamics equilibrium. This issue expresses that temperature of the universe outside and inside of the apparent horizon are similar.

\section{Conclusion}\label{VI}
In this paper, we have studied the $f(T, B)$ gravity with viscous fluid by flat-FRW background, in which $T$ and $B$ are the torsion scalar and the boundary term, respectively. It is interesting to know that the $f(T, B)$ model is described by one of the models of the $f(T)$ gravity and the $f(R)$ gravity along, by conversion relationship \eqref{rtb}, so, this model has a special value in cosmology. In what follows, we obtain the Friedmann equations in terms of function $f(T, B)$ and its partial derivatives, and then the corresponding Friedmann equations have been compared with its standard form. We have separately written the continuity equations in terms of universe components, namely, matter and dark energy. We note that there is an energy flow between the components of matter and dark energy by interaction term as $Q= 3 b^2 H \rho$, i.e., if dark energy flows to matter, then $Q >  0$ and vice versa.

In order to solve the model, we reconstructed the relationships of energy density, the pressure and the EoS of dark energy in terms of redshift parameter, then, we have taken the arbitrary function $f(T,B)$ as a power-law like function \eqref{ftb1}. In what follows, by fitting of parametrization function \eqref{rz1} with $38$ supernova data, we obtained the Hubble parameter function with respect to redshift parameter which the result has been shown in Fig. \ref{fig1}. Also, we have drawn the dark energy cosmological parameters with respect to redshift in the form the Figs. \ref{fig2} and \ref{fig3}. These figures showed us that the universe is undergoing accelerated expansion, so that the value of  EoS parameter is equal to $-1.09$ in late time evolution and is compatible with observational data.

Next, we analyzed the stability condition for the model by using of sound speed parameter. In that case, there is the stability condition in late universe, because the value of sound speed is bigger than zero in current time evolution so Fig. \ref{fig4} shows it. In the final section, we investigated  the generalized second law of thermodynamics for the viscous f(T,B) model in a flat--FRW universe with apparent horizon. By applying the thermodynamic equilibrium condition, the generalized second law of thermodynamics  is generality valid in the whole universe.

\end{document}